\documentstyle[aps,twocolumn]{revtex}

\draft

\begin{document}

\draft

\title{\bf Quantum probabilities as Bayesian probabilities}

\author{Carlton M. Caves,$^{(1)}$\cite{Permaddress} Christopher A.
Fuchs,$^{(1)}$ and R\"udiger Schack$^{(2)}$\medskip}

\address{$^1$Bell Labs, Lucent Technologies, 600--700
Mountain Avenue, Murray Hill, New Jersey 07974, USA
\\
$^2$Department of Mathematics, Royal Holloway, University of London,
Egham, Surrey TW20$\;$0EX, United Kingdom
\\
\medskip
}

\date{23 September 2001}

\maketitle

\begin{abstract}
  In the Bayesian approach to probability theory, probability
  quantifies a degree of belief for a single trial, without any {\it
  a priori\/} connection to limiting frequencies.  In this
  paper we show that, despite being
  prescribed by a fundamental law, probabilities for individual
  quantum systems can be understood within the Bayesian approach.  We
  argue that the distinction between classical and quantum
  probabilities lies not in their definition, but in the nature of
  the information they encode. In the classical world, {\it
  maximal\/} information about a physical system is {\it complete\/}
  in the sense of providing definite answers for all possible
  questions that can be asked of the system.  In the quantum world,
  {\it maximal information is not complete and cannot be completed}.
  Using this distinction, we show that any Bayesian probability
  assignment in quantum mechanics must have the form of the quantum
  probability rule, that maximal information about a quantum system
  leads to a unique quantum-state assignment, and that quantum theory
  provides a stronger connection between probability and measured
  frequency than can be justified classically.  Finally we give a
  Bayesian formulation of quantum-state tomography.
\end{abstract}

\medskip

\section{Introduction}

There are excellent reasons for interpreting quantum states as states
of knowledge \cite{FuchsPeres00}. A classic argument goes back to
Einstein \cite{Einstein}. Take two spatially separated systems $A$
and $B$ prepared in some entangled quantum state
$|\psi^{\scriptscriptstyle AB}\rangle$.  By performing the
measurement of one or another of two observables on system $A$
alone, one can {\it immediately\/} write down a new state for system
$B$---either a state drawn from a set $\{|\phi_i^{\scriptscriptstyle
B}\rangle\}$ or a set $\{|\eta_i^{\scriptscriptstyle B}\rangle\}$,
depending upon which observable is measured.  Since this holds no
matter how far apart the two systems are, Einstein concluded that
quantum states cannot be ``real states of affairs.''  For whatever
the real, objective state of affairs at $B$ is, it should not depend
upon the measurements made at $A$.  If one accepts this conclusion,
one is forced to admit that the new state (either a
$|\phi_i^{\scriptscriptstyle B}\rangle$ or a
$|\eta_i^{\scriptscriptstyle B}\rangle$) represents partial knowledge
about system $B$.  In making a measurement on $A$, one learns
something about $B$; the state itself cannot be construed to be more
than a reflection of the new knowledge.

The physical basis of
Einstein's argument has recently become amenable to experimental
test. Zbinden {\it et al.}~\cite{Zbinden2001} have reported an
experiment with entangled photons in which the detectors at $A$ and
$B$ are in relative motion. The experimental data rule out a certain
class of realistic collapse models, i.e., models in which the real
state of affairs at $B$ changes as a result of the measurement at
$A$. They also put a lower bound of $10^7$ times the speed of light
on the speed of any hypothetical quantum influence of the measurement
at $A$ on the real state of affairs at $B$.

We accept the conclusion of Einstein's argument and start from the
premise that ``quantum states are states of knowledge.''  An
immediate consequence of this premise is that all the probabilities
derived from a quantum state, even a pure quantum state, depend on a
state of knowledge; they are subjective or {\it Bayesian\/}
probabilities.  We outline in this paper a general framework for
interpreting all quantum probabilities as subjective.

If two scientists have different states of knowledge about a system,
they will assign different quantum states, and hence they will
assign different probabilities to the outcomes of some measurements.
This situation is commonly encountered in quantum cryptographic
protocols \cite{Gisin01}, where the different players, possibly
including an eavesdropper, have different information about the
quantum systems they are handling. In a Bayesian framework, the
probabilities assigned by the different players are all treated on an
equal footing; they are all equally valid. Subjective probability has,
therefore, no {\it a priori\/} connection to measured frequencies and
applies naturally to single quantum systems.

The Bayesian approach has been very successful in
statistics~\cite{Savage72,BernardoSmith,Science}, observational
astronomy~\cite{Loredo}, artificial intelligence~\cite{Microsoft},
and classical statistical mechanics~\cite{Jaynes83}. It seems to be
the general opinion, however, that the Bayesian interpretation is not
suitable for quantum-mechanical probabilities.  The probabilities
that come from a pure state are intrinsic and unavoidable.  How can
they not be objective properties when they are prescribed by physical
law?  How can Bayesian quantum state assignments be anything but
arbitrary?  Hasn't the tight connection between probability and
measured frequencies been verified in countless experiments?  What is
an experimenter doing in quantum-state tomography
\cite{Leonhardt,Buzek98} if not determining the unknown objective
quantum state of a system?

In this paper we give answers to these questions. These answers turn
out to be simple and straightforward. After a brief introduction to
Bayesian probability theory, we use Gleason's theorem
\cite{Gleason57} to show that any subjective quantum probability
assignment must have the form of the standard quantum probability
rule.  We then use a version of the so-called Dutch-book argument
\cite{DeFinettiBook,Earman} to show that if a scientist has {\it
maximal information\/} about a quantum system, he must assign a
unique pure state.  Our next step is to show that in the
case of maximal information, there is a conceptually simple
connection between (subjective) probability and (measured) frequency,
which is tighter than can be justified classically.  Finally, we
consider quantum-state tomography, where an experimenter is said to
be determining the ``unknown quantum state'' of a system from the
results of repeated measurements on many copies of the system.  An
``unknown quantum state'' is an oxymoron if quantum states are states
of knowledge, and we show how it can be eliminated from the
description of tomography by using a quantum version of the de
Finetti representation theorem for exchangeable sequences
\cite{Hudson76,Caves01}.  We conclude with a brief summary and an
outlook.

\section{Bayesian probability and the Dutch book}

Bayesian probabilities are degrees of belief or
uncertainty~\cite{BernardoSmith}, which are given an operational
definition in decision theory~\cite{Savage72}, i.e., the theory of
how to decide in the face of uncertainty.  The Bayesian approach
captures naturally the notion that probabilities can change when new
information is obtained.  The fundamental Bayesian probability
assignment is to a single system or a single realization of an
experiment.  Bayesian probabilities are defined without any
reference to the limiting frequency of outcomes in repeated
experiments.  Bayesian probability theory does allow one to make
(probabilistic) predictions of frequencies, and frequencies in past
experiments provide valuable information for updating the
probabilities assigned to future trials. Despite this connection,
probabilities and frequencies are strictly separate concepts.

The simplest operational definition of Bayesian probabilities is in
terms of consistent betting behavior, which is decision theory in a
nutshell.  Consider a bookie who offers a bet on the occurrence of
outcome $E$ in some situation.  The bettor pays in an amount
$px$---the {\it stake\/}---up front. The bookie pays out an amount
$x$---the {\it payoff\/}---if $E$ occurs and nothing otherwise.
Conventionally this is said to be a bet at {\it odds\/} of $(1-p)/p$
to 1.  For the bettor to assign a probability $p$ to outcome~$E$
means that he is willing to accept a bet at these odds with an
arbitrary payoff $x$ determined by the bookie.  The payoff can be
positive or negative, meaning that the bettor is willing to accept
either side of the bet. We call a probability assignment to the
outcomes of a betting situation {\it inconsistent\/} if it forces the
bettor to accept bets in which he incurs a sure loss; i.e., he loses
for every possible outcome.  A probability assignment will be called
{\it consistent\/} if it is not inconsistent in this sense.

Remarkably, consistency alone implies that the bettor must obey the
standard probability rules in his probability assignment: (i)
$p\ge0$, (ii) $p(A\vee B)=p(A)+p(B)$ if $A$ and $B$ are mutually
exclusive, (iii) $p(A\wedge B)=p(A|B)p(B)$, and (iv) $p(A)=1$ if $A$
is certain. Any probability assignment that violates one of these
rules can be shown to be inconsistent in the above sense.  This is
the so-called {\it Dutch-book argument}~\cite{DeFinettiBook,Earman}.
We stress that it does not invoke expectation values or averages in
repeated bets; the bettor who violates the probability rules suffers
a sure loss in a single instance of the betting situation.

For instance, to show that $p(A\vee B)=p(A)+p(B)$ if $A$ and $B$ are
mutually exclusive, assume that the bettor assigns probabilities
$p_A$, $p_B$, and $p_C$ to the three outcomes $A$, $B$, and $C=A\vee
B$. This means he will accept the following three bets: a bet on $A$
with payoff $x_A$, which means the stake is $p_Ax_A$; a bet on $B$
with payoff $x_B$ and thus with stake $p_Bx_B$; and a bet on $C$ with
payoff $x_C$ and thus with stake $p_Cx_C$ . The net amount the
bettor receives is
\begin{equation}
 R  = \cases{
x_A(1-p_A) -x_Bp_B + x_C(1-p_C)          & if $A\wedge\neg B$
\cr -x_Ap_A + x_B(1-p_B) + x_C(1-p_C)    & if $\neg A\wedge B$
\cr -x_Ap_A -x_Bp_B -x_Cp_C              & if $\neg A\wedge\neg B$}
\end{equation}
The outcome $A\wedge B$ does not occur since $A$ and $B$ are mutually
exclusive. The bookie can choose values $x_A$, $x_B$, and $x_C$
that lead to $R<0$ in all three cases unless
\begin{equation}
0=\det\pmatrix{
1-p_A &  -p_B &  1-p_C \cr
 -p_A & 1-p_B &  1-p_C \cr
 -p_A &  -p_B &   -p_C }
= p_A+p_B-p_C \;.
\end{equation}
The probability assignment is thus inconsistent unless
$p(A\vee B)=p_C=p_A+p_B$.

In our experience physicists find it difficult first to accept and
then to embrace the notion that subjective probabilities receive
their {\it only\/} operational significance from decision theory, the
simplest example of which is the Dutch-book argument in which
probabilities are {\it defined\/} to be betting odds.  In the
Dutch-book approach the structure of probability theory follows
solely from the requirement of consistent betting behavior.  There is
no other input to the theory.  For example, normalization of the
probabilities for exclusive and exhaustive alternatives is not an
independent assumption, so obvious that it needs no justification.
Instead normalization follows from probability rules~(ii) and (iv)
above and thus receives its sole justification from the requirement
of consistent betting behavior.

The only case in which consistency alone leads to a particular
numerical probability is the case of certainty, or {\it maximal
information}.  If the bettor is certain that the outcome $E$ will
occur, the probability assignment $p<1$ means he is willing to take
the side of the bookie in a bet on $E$, receiving an amount $px$ up
front and paying out $x$ if $E$ occurs, leading to a certain loss of
$x(1-p)>0$.  Consistency thus requires that the bettor assign
probability $p=1$.  More generally, consistency requires a particular
probability assignment only in the case of maximal information, which
classically always means $p=1$ or 0.

The quantum situation is radically different, since in quantum theory
maximal information is not complete~\cite{Israel}.  This
notwithstanding, we show that consistency still requires particular
probability assignments in the case of maximal information and, what
is more, that these probabilities are numerically equal to expected
limiting frequencies.  The keys to these results are Gleason's
theorem and a quantum variant of the Dutch-book argument of the
previous paragraph.

\section{Gleason's theorem and the quantum probability rule}

In order to derive the quantum probability rule, we make the
following assumptions about a quantum system that is described by a
$D$-dimensional Hilbert space:

\begin{itemize}
\item[(i)]Each set of orthogonal one-dimensional projectors,
$\hat\Pi_k=|\psi_k\rangle\langle\psi_k|$, $k=1,\ldots,D$ (the vectors
$|\psi_k\rangle$ make up an orthonormal basis), corresponds to the
complete set of mutually exclusive outcomes of some measurement,
i.e., answers to some question that can be posed to the system.
Throughout this paper, what we mean by a ``quantum question'' is a
measurement described by such a complete set of orthogonal
one-dimensional projectors.

\item[(ii)]The probabilities assigned to the outcomes are
consistent in the Dutch-book sense given above.

\item[(iii)]The probability assignment is {\it noncontextual\/}
\cite{Barnum}; i.e., the probability for obtaining the outcome
corresponding to a projector $\hat\Pi$ depends only on $\hat\Pi$
itself, not on the other vectors in the orthogonal set defining a
particular measurement.  As a consequence, it can be denoted
$p(\hat\Pi)$.
\end{itemize}

Condition~(ii) implies that, for each set of orthogonal
one-dimensional projectors,
\begin{equation}
\sum_{k=1}^D p(\hat\Pi_k) = 1 \;.
\end{equation}
Of course, this is simply the normalization condition, but in the
Bayesian view, normalization is enforced only by the requirement of
consistent betting behavior.  Except in the special case of a
two-dimensional Hilbert space, condition~(iii) then implies that
there exists a density operator $\hat\rho$ such that for every
projector $\hat\Pi=|\psi\rangle\langle\psi|$,
\begin{equation}
p(\hat\Pi)={\rm tr}(\hat\rho\,\hat\Pi)=\langle\psi|\hat\rho|\psi\rangle \;.
\label{eq:gleason}
\end{equation}
This is {\it Gleason's theorem}~\cite{Gleason57}. It means that,
under the assumptions of (i)~the Hilbert-space structure of quantum
questions, (ii)~Dutch-book consistency, and (iii)~probabilities
reflecting the Hilbert-space structure, any subjective probability
assignment must have the form~(\ref{eq:gleason}), which is the
standard quantum rule for probabilities.  Hence Bayesian ``degrees of
belief'' are restricted by the laws of nature, and any subjective
state of knowledge about a quantum system can be summarized in a
density operator $\hat\rho$.  Since one of the chief challenges of
Bayesianism is the search for methods to translate information into
probability assignments, {\it Gleason's theorem can be regarded as
the greatest triumph of Bayesian reasoning}.

\section{Maximal information and unique state assignment} \label{sec:unique}

Our concern now is to show that if a scientist has maximal
information about a quantum system, Dutch-book consistency forces him
to assign a unique pure state.  Maximal information in the classical
case means knowing the outcome of all questions with certainty.
Gleason's theorem forbids such all-encompassing certainty in quantum
theory.  Maximal information in quantum theory instead corresponds to
knowing the answer to a maximal number of questions (i.e.,
measurements described by one-dimensional orthogonal projectors).
Suppose then that a scientist is certain about the outcome of all
questions that share one particular projector
$\hat\Pi=|\psi\rangle\langle\psi|$. The scientist is certain that the
outcome corresponding to this projector will occur in response to any
of these questions, so Dutch-book consistency requires that its
probability be $p=1$.  Now let $\hat\rho$ be the state assigned to
the system.  In the language of Gleason's theorem, we have
$\langle\psi|\hat\rho|\psi\rangle=1$. This implies that
$\hat\rho=\hat\Pi=|\psi\rangle\langle\psi|$. Gleason's theorem
further implies that the scientist cannot be certain about the
outcome of any other questions, so this is the case where he has
maximal information.  Maximal information thus leads to the
assignment of a unique pure state.

Given the assumptions of Gleason's theorem, if a scientist has
maximal information, any state assignment that is different from the
unique pure state derived in the last paragraph is inconsistent in
the Dutch-book sense; i.e., it leads to a sure loss for a bet on the
outcome of a measurement on a single system that includes the unique
pure state among the outcomes.  The Hilbert-space structure of
quantum questions plus noncontextuality alone puts this tight
constraint on probability assignments.

We emphasize that the uniqueness of the quantum state assignment
holds even though no measurement allows an experimenter to decide
with certainty between two nonorthogonal pure-state assignments.
Though maximal information leads to a unique pure state, the state
assignment cannot be verified by addressing questions to the system.
Finding out the state assignment requires consulting the assigner or
the records he leaves behind.  This property is another reason for
regarding quantum states as states of knowledge.

In both the classical and the quantum case, consistency enforces a
particular probability assignment if and only if there is maximal
information.  In the classical case, maximal information corresponds
to certainty, i.e., the trivial probability assignment 1 or 0, so
classically maximal information is {\it complete}.  In quantum
mechanics, maximal information leads to a unique pure state
assignment $|\psi\rangle\langle\psi|$, which is equivalent to
prescribing (generally nontrivial) probabilities for all possible
measurements. In quantum mechanics, {\it maximal information is not
complete and cannot be completed}.

\section{Subjective probability and measured frequency}

Up to this point, we have not mentioned repeated experiments or
long-run frequencies. Both the Dutch-book argument and Gleason's
theorem are formulated for single systems.  There is no
justification, at this point, for identifying the probabilities
derived from Gleason's theorem with limiting frequencies.  To make
the connection between the above results and repeated measurements,
an additional assumption is needed, namely that the Hilbert space of
$N$ copies of a quantum system is given by the $N$-fold tensor
product of the single-system Hilbert space.  In doing so, we are
assuming that the $N$ copies of the quantum system are labeled by
some additional degree of freedom that renders irrelevant the
symmetries required for identical particles.

Now assume that a scientist has maximal information about $N$
copies of a quantum system, specifically the same maximal information
about each system.  Applying Gleason's theorem and the Dutch-book
argument of Sec.~\ref{sec:unique} to the tensor-product Hilbert space
leads to a unique pure product-state assignment
$\hat\rho^{(N)}=\hat\Pi\otimes\cdots\otimes\hat\Pi$, where
$\hat\Pi=|\psi\rangle\langle\psi|$. Suppose that repeated
measurements are performed using the single-system projectors
$\hat\Pi_k=|\psi_k\rangle\langle\psi_k|$, $k=1,\ldots,D$. The
probability of obtaining the sequence of outcomes $k_1,\ldots,k_N$ is
given by
\begin{equation}
p(k_1,\ldots,k_N)=
{\rm tr}
\Bigl(\hat\rho^{(N)}\hat\Pi_{k_1}\otimes\cdots\otimes\hat\Pi_{k_N}\Bigr)=
p_{k_1}\cdots p_{k_N}\;,
\label{eq:pk}
\end{equation}
where
\begin{equation}
p_k={\rm tr}(\hat\Pi\,\hat\Pi_k)=|\langle\psi_k|\psi\rangle|^2\;.
\end{equation}
This means that the outcomes of repeated measurements are {\it
independent and identically distributed\/} (i.i.d.).  The probability
for outcome $k$ to occur $n_k$ times, where $k=1,\ldots,D$ and
$\sum_k n_k=N$, is given by the multinomial distribution,
\begin{equation}
\label{iid}
p(n_1,\ldots,n_D)={N!\over n_1!\cdots n_D!}\,p_1^{n_1}\cdots p_D^{n_D}\;,
\end{equation}
which peaks for large $N$ at $n_k\simeq Np_k$, $k=1,\ldots,D$.  The
probability of observing frequencies $n_k/N$ close to $p_k$ converges
to 1 as $N$ tends to infinity.

In the classical case an i.i.d.\ assignment is often the starting
point of a probabilistic argument.  Yet in Bayesian probability
theory, an i.i.d.\ can never be strictly justified except in the case
of maximal information, which in the classical case implies certainty
and hence trivial probabilities.  The reason is that the only way to
be sure all the trials are identical in the classical case is to know
everything about them, which implies that the results of all trials
can be predicted with certainty \cite{Jaynespost}.   In contrast, to
ensure that all systems are the same in quantum mechanics, it is
sufficient to have the maximal, but incomplete information that leads
to a unique pure state.  Thus the quantum i.i.d.\
assignment~(\ref{iid}) is a consequence of Dutch-book consistency and
the Hilbert-space structure of quantum mechanics.

To summarize, in quantum mechanics maximal information leads to
nontrivial i.i.d.\ assignments.  Maximal information means that the
pure product-state assignment is the unique consistent state
assignment.  From the pure product-state assignment comes the i.i.d.\
for the outcomes of any repeated measurement.  Together with
elementary combinatorics, this gives the strict connection between
probabilities and frequencies displayed in the laws of large numbers.
In this sense, the equality between probability and limiting
frequency holds only in quantum mechanics.

\section{Unknown quantum states and the quantum de Finetti
representation}

An important practical use of repeated measurements on many copies
of a quantum system is in {\it quantum-state
tomography\/}~\cite{Leonhardt,Buzek98}.  The data gathered from the
measurements is said to determine the ``unknown quantum state'' of
the system. But what can an unknown quantum state mean?  If a quantum
state is a state of knowledge, then it must be known by somebody.  If
the Bayesian interpretation of quantum probabilities is to be taken
seriously, there must be a way to eliminate the ``unknown quantum
state'' from the description.

The key to this excision is to identify the salient feature of the
state of knowledge that applies when an experimenter performs
quantum-state tomography.  That salient feature is that the
experimenter can contemplate examining an arbitrarily large number of
systems, all of which are equivalent from his perspective.  This
means (i)~that the density operator $\hat\rho^{(N)}$ for $N$ systems
should be {\it symmetric}, i.e., invariant under all permutations of
the $N$ systems, and (ii)~that this symmetry should hold for all
values of $N$, with the consistency requirement that
$\hat\rho^{(N)}$ arises from tracing out one of the systems in
$\hat\rho^{(N+1)}$.  A sequence of density operators that satisfies
these two properties is said to be {\it exchangeable}, by analogy
with de Finetti's definition~\cite{DeFinettiBook} of exchangeable
multi-trial probabilities.

The quantum de Finetti representation theorem~\cite{Hudson76,Caves01}
establishes that for any exchangeable sequence of density operators,
$\hat\rho^{(N)}$ can be written {\it uniquely\/} in the form
\begin{equation}
\hat\rho^{(N)}=\int d\rho\,p(\rho)\,\rho\otimes\cdots\otimes\rho\;,
\end{equation}
where the tensor product includes $N$ terms, the integral runs over
the space of density operators, and the ``generating function''
$p(\rho)$ can be thought of as a normalized ``probability density on
density operators.''  An exchangeable density operator captures what
an experimenter knows about the systems he intends to examine.  It is
a primary quantum-state assignment for multiple copies, with no
mention of unknown quantum states.  The content of the quantum de
Finetti representation is that the exchangeable state assignment can
nevertheless be thought of in terms of unknown density operators;
ignorance of which density operator is described by the generating
function.

Exchangeability permits us to describe what is going on in
quantum-state tomography.  Suppose two scientists make different
exchangeable state assignments and then jointly collect data from
repeated measurements.  Suppose further that the measurements are
``tomographically complete''; i.e., the measurement probabilities for
any density operator are sufficient to determine that density
operator.  The two scientists can use the data $D$ from an initial
set of measurements to update their state assignments for further
systems.  In the limit of a large number of initial measurements,
they will come to agreement on a particular product state
$\hat\rho_{\scriptscriptstyle D}\otimes\hat\rho_{\scriptscriptstyle
D}\otimes\cdots$ for further systems, where
$\hat\rho_{\scriptscriptstyle D}$ is determined by the data. {\it
This is what quantum-state tomography is all about.}  The updating
can be cast as an application of Bayes's rule to updating the
generating function in light of the data \cite{Schack01}.  The only
requirement for ``coming to agreement'' is that both scientists
should have allowed for the possibility of
$\hat\rho_{\scriptscriptstyle D}$ by giving it nonzero support in
their initial generating functions.

\section{Summary and outlook}

We promised simple answers, and it's hard to imagine simpler ones.
The physical law that prescribes quantum probabilities is indeed
fundamental, but the reason is that it is a fundamental rule of
inference---{\it a law of thought}---for Bayesian probabilities.  It
follows from requiring Dutch-book consistency for probability
assignments that are faithful to the Hilbert-space structure of
quantum questions.  These same desiderata require a particular
pure-state assignment when a scientist has maximal information, and
because maximal information is not complete, they give a strict
connection between observed frequencies and pure-state quantum
probabilities.  The notion of an ``unknown quantum state,''
irreconcilable with the idea of quantum states as states of
knowledge, can be banished from quantum-state tomography using the
quantum de Finetti representation.

Quantum information science~\cite{NatureReview} is an emerging field
that uses quantum states to escape the constraints imposed on
information processing in a realistic/determin\-is\-tic world.  The
rewards in quantum information science are great: teleportation of
quantum states, distribution of secret keys for encoding messages
securely, and computations done more efficiently on a quantum
computer than on any classical machine.  The key to these rewards is
that a quantum world is less constrained than a classical one.  As
quantum information science harnesses the greater range of
possibilities available in the quantum world, we believe it is
imperative to understand and elucidate the fundamental principles
underlying quantum mechanics. In this paper we show how to interpret
quantum states consistently as states of knowledge, reflecting what
we know about a quantum system.  This is just one step in a broader
program to try to disentangle the subjective and objective aspects of
the quantum world \cite{Fuchs01}.  We leave the last word to Edwin
T.~Jaynes \cite{Jaynes1990a}, who inspired us to pursue the Bayesian
view:

\begin{quotation}
Today we are beginning to realize how much of all physical science is
really only {\it information}, organized in a particular way.  But we
are far from unravelling the knotty question: ``To what extent does
this information reside in us, and to what extent is it a property of
Nature?''\,\dots\ Our present QM formalism is a peculiar mixture
describing in part laws of Nature, in part incomplete human
information about Nature---all scrambled up together by Bohr into an
omelette that nobody has seen how to unscramble.  Yet we think the
unscrambling is a prerequisite for any further advance in basic
physical theory \dots\,.
\end{quotation}

\section{Acknowledgments}

CMC was supported in part by U.S.\ Office of Naval Research Grant
No.~N00014-93-1-0116.

\end{document}